
\magnification\magstep1
\hsize=6truein

\def\go{
\mathrel{\raise.3ex\hbox{$>$}\mkern-14mu\lower0.6ex\hbox{$\sim$}}
}
\def\lo{
\mathrel{\raise.3ex\hbox{$<$}\mkern-14mu\lower0.6ex\hbox{$\sim$}}
}
\def\title#1{\centerline{\bf #1}}
\def\author#1{\bigskip\bigskip{\centerline{#1}}\bigskip}
\def\address#1{\bigskip{\centerline{#1}}\bigskip\bigskip}
\def\sec#1{\bigskip{\centerline{\bf#1}}\medskip}
\def\subsec#1{\medskip\noindent{\it #1}\medskip}
\newcount\secno
\newcount\eqnum

\def\equno{
\global\advance\eqnum by1\eqno({\the\secno.\the\eqnum})
}



\def\enam#1{
\xdef#1{(\the\secno.\the\eqnum)}
}

\title{ON THE JETS ASSOCIATED WITH GALACTIC SUPERLUMINAL SOURCES}
\author{Amir Levinson and Roger Blandford}
\address{Theoretical Astrophysics 130-33, Caltech, Pasadena CA 91125.}

\sec{Abstract}
Recent observations of GRS 1915+105 and GRO J1655+40 reveal
superluminal motions in Galactic sources.  This letter examines the
physical conditions within these Galactic sources, their
interaction with their environment, their possible formation, and
contrasts them with their extragalactic counterparts.  In particular,
e$^{+}$-e$^{-}$ and e-p jets are contrasted, constraints on particle
acceleration in the jets are imposed
using X-ray and radio observations, the $\gamma$-ray flux from
e$^+$-e$^-$ jets expected at EGRET energies and the flux in infrared
lines from an e-p jet are estimated.  It is also suggested that these
sources may exhibit low frequency radio lobes extending up
to several hundred parsecs in size, strong, soft X-ray absorption
during the birth of the radio components and emission line strengths
anti-correlated with the X-ray flux. The implications for other X-ray
transients are briefly discussed.

\sec{1. Introduction}
There is now direct evidence for superluminal motion in the radio images
for two strong
Galactic X-ray transient sources, GRS 1915+105 and GRO J1655+40 (Mirabel \&
Rodriguez 1994, hereafter MR94; Tingay et al. 1995; Hjellming \& Rupen
1995, hereafter HR95 ).
These motions are probably associated with
relativistic jets emanating from a black hole in a X-ray binary.
Several other transient radio sources associated
with soft X-ray novae may also involve collimated jets (HR95).  In
this letter we use existing radio, optical and X-ray
observations to place constraints on the physical conditions within
these radio-emitting X-ray transients, contrasting
them with their extragalactic counterparts.
We first analyze the jets in some generality and  then
speculate upon how they may be collimated.  We then illustrate these ideas
using the two known examples and finally list possible observations that
may further elucidate the natures of these sources.

\sec{2. Relativistic Jets}
\subsec{2.1 Synchrotron emission}
The radio emission is observed to originate in pairs of radio components
that move away from the central source with mildly relativistic speed.
The particle acceleration and magnetic field amplification responsible
for the synchrotron emission may both be caused by internal
shocks propagating along a jet. Consider one such jet formed at
$r\equiv10^xr_x$~cm
$\sim10^6$~cm near a black hole in a X-ray binary of orbital radius
$a\sim10^{12}$~cm, that propagate out into interstellar space $r\lo10^{20}$~cm
with an opening angle $\phi(r)$.
Relativistic electrons accelerated {\it in situ} should radiate by the
synchrotron and inverse Compton processes and the former should dominate
at large radius.  From the resolved radio images, it is possible
to estimate a fiducial equipartition field strength $B^\ast$,
in the normal manner using the time-averaged
radio intensity and assuming (falsely) that the source is stationary.
(Numerically, $B^\ast\sim4(T_{B6}/\phi_{-1}r_{16})^{2/7}\nu_9^{5/7}$~mG, where
the brightness temperature $T_B=10^6T_{B6}$~K is
evaluated at the frequency $\nu=\nu_9$~GHz assuming that
the radio spectral index $\alpha_R\sim0.5$, e.g., Rybicki \& Lightman
1979.)

If the jet Lorentz factor is $\Gamma$
and its velocity makes an angle $\theta$
with the line of sight, we can define an observer's Doppler
factor $\delta_o=[\Gamma(1-\beta\cos\theta)]^{-1}$.  As the synchrotron
cooling time is long compared with the dynamical time, the jet
power associated
with the emitting electrons and the electromagnetic field satisfies
$$
L_{je}\go{c\over2}\left({\Gamma B^\ast\phi r\over\delta_o^{5/7}}\right)^2
\sim10^{34}\Gamma^2\delta^{-10/7}_o(B^*_{-3})^2\phi^2_{-1}
r^2_{16}{\rm erg s}^{-1}
\eqno(1)
$$
where rough equality occurs at equipartition and the jet power rises
$\propto(P_{mag}/P_e)^{-3/4}$ if the jet is particle-dominated and $\propto
(P_{mag}/P_e)$, if it is magnetically-dominated.

The inertia of the surrounding gas will decelerate the head of
the jet.  The speed of advance of the head $V_h$ can be estimated by
equating the ram pressure of
the surrounding material measured in a frame comoving with the head,
with the jet thrust (e.g, Begelman et al. 1984).
As long as $V_h$ is supersonic, strong
shocks may be formed at the outer lobes, as in extragalactic FRII
sources, and can give rise to enhanced,
steep spectrum synchrotron emission that can be seen at low radio
frequencies as hot spots.  If $V_h$ is subsonic, then low frequency radio
observations may reveal a source similar to FRI objects.

To estimate the maximum radio power emitted by the lobes, we assume
that the lobes subtend $\sim1$ steradian so that their
pressure is $\sim L_j^{2/3}\rho^{1/3}r^{-4/3}\sim3\times
10^{-7}L_{j38}^{2/3}n^{1/3}r_{18}^{-4/3}$ dyne cm$^{-2}$. (In contrast
to the powerful extragalactic sources, radiation loss is unlikely to
be important in the hot spots.)  The lobe brightness temperature at
a fiducial frequency $\sim$ 5 GHz then satisfies
$$
T_{B5}\lo4\times10^5L_{j38}^{7/6}n^{1/2}r_{18}^{-4/3}\ \ \ {\rm K}
\eqno(2)
$$
where equality requires equipartition.  If the source is too young or
particle acceleration is too inefficient, the brightness temperature
will be much less than the upper limit.  Otherwise, these lobes should
be detectable against the normal Galactic radio backgound.  If there
is a small population
of ultrarelativistic Galactic jets then it is possible that none of
them will be beamed
towards us and that we can only detect their presence from their double radio
lobes.

\subsec{2.2 Inverse Compton scattering}

Accelerated relativistic electrons can also radiate by
inverse Compton scattering of accretion disk radiation. Let the
energy density be dominated by photons of energy $E_X\sim1$~keV
(in contrast to $\sim10$~eV for extragalactic sources)
and luminosity $L_s(r)$ and let the
characteristic Doppler factor for transforming this radiation into the
frame of the jet be $\delta_j(r)$ so that the associated photon energy
in the jet frame is $E_X'\sim\delta_jE_X$.
Introduce a characteristic electron cooling energy in the jet frame
by equating the radiative cooling time to the outflow timescale
$$
E'_{ec}(r)\sim3\left({\Gamma\over\beta k_j}\right)
L_{s38}^{-1}r_{12}{\rm GeV},
\eqno(3)
$$
where $k_j\sim<\delta_j^2>\sim\Gamma^2r_7^{-4}$ for direct
illumination by the accretion disk
and $k_j\sim\Gamma^2\tau$, if this radiation is scattered locally by free
electrons in the surrounding medium with local Thomson depth $\tau$.
Inverting Eq.~(3), we can define a cooling radius $r_c(E'_e)$, within
which electrons of energy $E'_e$ will cool.
In order to accelerate an electron to an energy $\go E'_{ec}$ requires
that the particle acceleration occur impulsively
on a timescale $t'_{acc}<r/\Gamma c$.  The maximum $\gamma$-ray energy that
can be scattered is then $\delta_o(r/\Gamma ct'_{acc})E'_{ec}$.  The scattered
flux and spectrum depend upon the fraction of the jet kinetic energy that is
transformed into relativistic particles in this manner.
If a fraction $\eta$ of the jet power is emitted as
$\gamma$-rays, the integrated $\gamma$-ray flux at Earth will be,
$$
F_{\gamma}\sim10^{-8}\eta\delta_o^2L_{j38}D_4^{-2}
{\rm ergs\ s^{-1}\ cm^{-2}}
\eqno(4)
$$
where $D_4$ is the distance to the source in 10$^4$ pc.
Further features of the Compton scattering depend upon whether the jet is pair
- or proton - dominated.
\subsec{2.3 e$^{\pm}$ jet}
If $\gamma$-rays are emitted at small enough radius, they will not be able to
escape without creating electron-positron pairs.  These, in turn, can
produce lower energy $\gamma$-rays and a cascade will develop which terminates
when the $\gamma$-ray has a low enough energy to escape.  The region from
which $\gamma$-rays of a given energy can escape is known as the
{\it $\gamma$-sphere} and its radius is
$$
r_{\gamma}(E_\gamma)\simeq3\times10^{7}k_{pp}L_{E44}(m_e^2c^4/E_\gamma)
\ \ \ \ {\rm
cm},
\eqno(5)
$$
where $10^{44}L_{E44}$ s$^{-1}$ is the spectral luminosity of the
central X-ray source,
$k_{pp}\sim<1-\cos\phi>$ for radiation from an accretion disk that
propagates at an angle $\phi$ to the jet and $k_{pp}\sim\tau$ if the local
scattered component dominates (c.f., Blandford \& Levinson 1995).
We can invert Eq.~(5) to define the the threshold energy $E_{\gamma th}(r)$
which is the maximum energy of an escaping $\gamma$-ray from radius $r$.

Now, pairs will cool to subrelativistic
energies for $r\lo r_c(m_ec^2)\sim3\times10^{8}k_jL_{s38}$\break $\Gamma^{-1}$
cm.
Their density will be limited
by annihilation (Blandford \& Levinson 1995).  We can also define an
{\it annihilation radius}, within which
the density of annihilated pairs becomes smaller than that required to
carry the jet power.  Consequently,
pair jets require the presence of some other
carrier of energy and momentum.  In the absence of baryons this is presumably
electromagnetic. For subrelativistic pairs, $r_{ann}\sim3\times10^9
L_{j38}\Gamma^{-1}$~cm, and is somewhat less if the pairs remain relativistic.

Another important difference
between the Galactic and the extragalactic \break sources
is that the former have much steeper X-ray spectra and consequently
their $\gamma$-spheric radii increase more rapidly with increasing
$\gamma$-ray energy.
Furthermore, $E_{th}(r_{ann})\sim1$~GeV, instead of $\sim1$~MeV as in
the extragalactic case.  This probably means that the $\gamma$-ray
spectrum will be flatter in the MeV-GeV
range.  Furthermore, the bulk Lorentz factors of the jets in the
bright EGRET AGN sources are typically of order 10, much larger than
those inferred for the jets in the Galactic sources.

A plausible picture of Galactic e$^{\pm}$ jets, based on the above
results, is as follows: some fraction of the accretion luminosity (or
the spin energy of a rotating black hole) is
extracted from the central source in the form
of a collimated electromagnetic jet.
During quiescent states, the jet is essentially
invisible.  However, when either the particle acceleration is sufficiently
rapid
or a reduced ambient radiation field renders the inverse Compton
radiation loss sufficiently ineffective,
pairs can be injected to energies above $E_{th}(r)$
and an intense pair cascade is initiated.  At
this radius a transition to a particle dominated flow occurs via the
evolution of the cascade, leading to $\gamma$-ray emission
and the eventual formation of a superluminal radio feature.
If the cascade is initiated within the annihilation
radius, the mildly relativistic pairs will be annihilated and
the radio spectrum will exhibit a low energy cutoff.
\subsec{2.4 e-p jet}
If the jet is accelerated and collimated close to the black hole
as an e-p plasma, perhaps through the agency of radiation pressure,
and particle acceleration is inefficient above $\go E_{\gamma
th}(r)/\Gamma$, then pairs are not created and
the minimum jet power is larger than that given by Eq.~(1) by a factor
$\sim[m_p/\gamma_{min}\ln(\gamma_{max}/\gamma_{min})m_e]^{4/7}$,
where the electron distribution function is supposed to extend from
$\gamma_{min}$ to $\gamma_{max}$.  Typically, this factor is $\sim3-30$.
(Alternatively, a pair jet may form as described above and plasma from
the surrounding wind may be entrained.)

One possible diagnostic of e-p jets is the presence of
Doppler-shifted spectral lines, such as $H_{\alpha}$, as seen in
SS433.
Due to the relativistic motion of the jet the line
will be Doppler shifted by the approaching and receding Doppler
factors $\delta_o$.
Following Begelman et al. (1980) and
Davidson \& McCray (1980), we suppose that the gas in the
line-emitting region is clumped, and denote by $\varepsilon$ the
volume filling factor of the dense blobs comprising the line-emitting
beam and by $10^{15}R_{15}$ cm the beam's length.  The
$H_{\alpha}$ emissivity should lie in the range between $10^{-24.6}$
and $10^{-23}$ ergs cm$^3$ s$^{-1}$ at $\sim10^4$ K (Davidson \&
McCray 1980), depending on the density in the emitting blobs.  Let us
adopt the value $10^{-23.5}$ ergs cm$^3$ s$^{-1}$.  We then obtain for
the emitted flux
$$
F_{H_{\alpha}}\simeq 10^{-23}\delta^2\left({L_{j38}\over
D_4}\right)^2 (\varepsilon R_{15}
\phi^2\Gamma^4)^{-1}\ \ {\rm ergs\ cm^{-2}
\ s^{-1}},
\eqno(6)
$$
where it has been assumed that the average density of the hot phase is
$\varepsilon n_{\rm cold}$.  The cold blobs should be confined by the
pressure of the hot phase
in the jet or, alternatively, by the magnetic fields.
\sec{3. Jet Formation and Confinement}
The Galactic superluminal sources further demonstrate that
relativistic
jet formation
can operate on a stellar as well as a galactic scale.  Presumably,
the common feature is the presence of an accretion disk orbitting in a
relativistically deep potential well.  In order
to explore how this might occur in a ($M\sim3-10$~M$_\odot$)
binary X-ray source,
we suppose that the jet is collimated by a wind emanating
from the disk surface over a range of
radii from $\lo10^7$ to $\go10^{11}$~cm with speed.
$V_W=1000V_{W8}$~km s$^{-1}$ declining with cylindrical radius.
As the jet propagates away from its source, there will be radial transport
of linear momentum which will flatten the velocity profile.  If most
of the momentum derives from large disk radius, then the asymptotic jet
speed will be $V_{W8}\sim1-10$.

It has long been argued in the case of AGN and protostellar jets, that a
hydromagnetic wind is a more plausible collimator than a purely hydrodynamic
wind because when the field is primarily toroidal, the
transverse force density is $-r_\perp^{-2}d(r_\perp^2P_{mag})/dr_\perp$
(as opposed to $-dP_{gas}/dr_\perp$ for gas pressure) allowing a smaller
magnetic pressure
to focus a jet of larger total pressure.  In addition,
$P_{mag}$ declines less rapidly than $P_{gas}$ as the wind expands
which implies that magnetic collimation is likely to become relatively
more important.
We adopt the magnetic collimation hypothesis,
though much of what follows is more general.

For $r<<a$, magnetic confinement of the jet
can be relatively effective with each nested magnetic surface
confining the interior flow, until ultimately the inertia of the wind from
the outer disk prevents transverse expansion. However, this magnetic
focusing cannot provide much pressure amplification after the jet has
propagated out to a radius comparable with the outer radius of the
disk.  At this point, either the jet itself must have sufficient
internal density to be effectively free and travel hypersonically with
Mach number $\go\phi^{-1}$ or it must be confined by the inertia of
the surrounding wind.  The former possiblity may be relevant for e-p
jets.
However, we suspect that pair jets require external confinement at
this radius. The forgoing considerations suggest
that the jet itself exerts a transverse pressure of
$$
P\sim3\times10^4L_{j38}\Gamma^{-2}\beta^{-1}
r_{12}^{-2}\phi_{-1}^{-2}{\rm dyne cm}^{-2}
\eqno(7)
$$
which will cause the surrounding, slower wind
to expand with a speed $\sim(P/\rho_{Wj})^{1/2}$\break$\lo V_{Wj}\phi$, where
$Wj$
denotes values of the wind density and speed averaged within a few jet widths.
The {\it minimum} wind discharge for ultimate
inertial confinement at radius $r\sim a$ is then given by
$$
\dot M_{Wj}\go10^{-4}L_{j38}\Gamma^{-2}\beta^{-1}\phi_{-1}^{-2}V_{Wj8}^{-1}
{\rm M}_\odot{\rm yr}^{-1}
\eqno(8)
$$
and the associated wind power is $\go10^{38}L_{j38}\Gamma^{-2}\beta^{-1}
\phi_{-1}^{-2}V_{Wj8}$~erg s$^{-1}$. (In making this estimate
we have supposed that most of the discharge is confined to a polar wind
with transverse scale $\sim3$ jet radii.  If the wind fills a larger
solid angle, the discharge and power must correspondingly be increased.)
This wind will propagate well beyond the observed radio sources before
terminating through a strong shock when its momentum flux balances
the ambient interstellar pressure.

We can now use this simple prescription to estimate the physical conditions
in the wind.  If we measure the wind discharge as
$\sim10^{-6}\dot M_{Wj-6}$~M$_\odot$~yr$^{-1}$, its Thomson optical
depth is likely to be
$\tau_T(r)\sim0.1\dot M_{Wj-6}V_{W8}^{-1}
r_{12}^{-1}$M$_\odot$~yr$^{-1}$, for $r_{12}\go0.1$.
For $r_{12}\lo0.1$, we emphasize that the optical depth need not
be much greater than this value because of the efficacy
of magnetic confinement. However, the wind that we postulate
is likely to
extinguish any soft X-ray flux if it is strong enough to collimate
the jets.

This wind may also be observable optically.
Its thermal state depends upon the photoionizing flux.
The ionizing parameter
is $U\sim0.1L_{UV36}V_{W8}/\dot M_{Wj-6}$.  For $0.1\lo U\lo10$, a two
phase medium is possible with hot Compton-heated gas at a temperature
$T\sim10^7-10^8$~K coexisting with line-emitting gas at a temperature
$T\sim 10^4$~K. (At the density envisaged, the thermal equilibration
time turns out to be short compared with outflow time.)  Now suppose
that the mass accretion rate increases as a consequence of some
disk instability.  The ultraviolet and X-ray emission will increase
as a consequence of enhanced dissipation at the inner disk.  This in
turn will heat the gas so that the pressure is largely thermal as
opposed to largely magnetic.  We propose that this prevents effective
magnetic collimation and consequently a jet does not form.  When the
disk accretion rate falls, the ionization parameter falls and the gas
in the wind cools so that it becomes magnetically dominated.  This
allows a collimating hydromagnetic wind to form.  If there is also
a central source of relativistic plasma or electromagnetic energy,
perhaps derived from the spin of a central black hole, then this will
form the radio-emitting core of the radio jet. This possibility is
relevant to the observation that the radio outbursts appear to follow
the X-ray outbursts in GRO~1655-40 (see below).
A possible, alternative model for
an accretion rate-radio jet connection has been proposed by Meier
(1995).
This wind is also likely to be a source of optical and ultraviolet
emission lines and our model predicts that the fraction of the
bolometric flux
reprocessed in the form of emission lines should be anticorrelated
with the X-ray flux.
\sec{4. Interpretation of GRS 1915+105, GRO J1655-40}
\subsec{4.1 GRS 1915+105}
This source is at a distance of $D_4\sim1.25$ (MR94), and exhibits
X-ray luminosity of a few times 10$^{38}$ ergs
s$^{-1}$ (Harmon et al. 1994).  It has been observed by the VLA at 5
and 10 GHz.  Following MR94 we assume, for simplicity,
that the pattern speed and the flow speed are equal (c.f., Bodo
\& Ghisellini 1995).  The inferred speed and angle to the line of
sight of the ejecta are then $\beta\sim0.92$ and $\theta= 70\pm2^{\circ}$
(MR94), corresponding to $\delta_o\simeq0.57$.  (This measurement
allows us to predict the wavelengths of, for example, possible
$H_\alpha$ lines, namely $\sim$ 1.15, 2.15 $\mu$.)
The radio features appear to move
away at a constant speed out to a distance of at least 0.1 pc from the
central source (MR94).

On March 24, 1994 the measured flux was $\sim0.7$~Jy.  Even
though the source was not resolved at that time, the inferred
distance from the putative core was $\sim$0.08 arcsec, and the blob
size was about 0.06 arcsec, corresponding to
a linear size of $\sim 10^{16}$ cm.
We estimate $P_{min}\simeq1.8\times10^{-5}$ dyn cm$^{-2}$, and
$B^\ast\simeq2\times10^{-2}$ G, at a distance of $\sim 10^{16}$ cm
from the central source.  Eq. (1) then gives $L_{j38}\go2$ for
$e^{\pm}$ jet, and about 4 times that for $e-p$ jet.
The annihilation radius $ r_{ann}\go2.4\times10^9$
cm.  Taking $\kappa_{pp}L_{s38}=10^{-2}$ yields
$ r_{\gamma}\simeq2.5\times10^8\left({E_{\gamma}/10^3}\right)^{0.5}$
cm.  Since $r_{\gamma}(1GeV)<r_{ann}$ low frequency cutoff of the
radio spectrum from the jet may be expected.  The acceleration time
required for a
formation of a particle loaded blob is $\lo10^{-3}(r/c)$.  From
equation (4) it follows that $F_{\gamma}\go3.5\times10^{-9}\eta$ ergs
cm$^{-2}$
s$^{-1}$.  If the radiative efficiency $\go0.1$, $\gamma$-ray
outbursts from this source, if comprises of e$^{\pm}$ pairs, might be
detectable.  From equation (6) we obtain $F_{H_{\alpha}}\lo10^{-14}$
ergs s$^{-1}$ cm$^{-2}$.

\subsec{4.2 GRO J1655-40}

For this source we adopt the parameters inferred by HR95, namely
$D_4=0.31$ (this distance is in a very good
agreement with the distance inferred by Bailyn et al. 1995, based on
interstellar absorption), $\beta=0.92$, and
$\delta_o\simeq0.46$.  (The predicted $H_\alpha$ wavelengths are
$1.4,1.8\mu$.)
The lowest frequency observed with the VLA was
$\nu_{10}=0.15$.  The light curves indicate a peak flux of 5.5 Jy at
this frequency 6 days after the beginning of the observations,
implying $l<10^{16}$ cm, and $T_{B5}\sim10^7$ K.  During the high X-ray
state Harmon et al. (1995)
derived a 20-100 keV luminosity of about 10$^{37}$ ergs s$^{-1}$ and
energy spectral index of $-$3.1.
The spectrum appears to harden to $-$2.5 when the flux drops.  If
we extrapolate the spectrum down to 1 keV, we estimate $L_{s38}\lo1$
during high states.  However, the X-ray luminosity was in fact smaller
during the peak of the radio flux, as discussed below.
For illustration we shall assume $\kappa_{pp}L_{s38}=10^{-3}$.
Adopting these parameters we obtain, $P_{min}\simeq5\times10^{-6}$ dyn
cm$^{-2}$, $B^\ast\simeq10^{-2}$ G, $L_{j38}>0.6$,
$r_{ann}>7\times10^8$ cm, and
$r_{\gamma}\sim10^{7}(E_{\gamma}/10^3)^{0.5}$ cm, for $e^{\pm}$ jet.
For e-p jet $L_{j38}\go3$.  The $H_{\alpha}$ flux obtained from Eq.
(6) is roughly the same as that obtained for GRS 1915

Bailyn \& Orosz (1995) have measured a spectroscopic orbital period of
2.6~d and a mass function $f_1=3.35\pm0.14$~M$_\odot$ implying that
the compact object is a black hole with a mass $5.3$~M$_\odot$\
and $a\sim1\times10^{12}$~cm.  They have also
reported observations of a hard optical
continuum with spectral index $d\ln F_\nu/d\ln\nu\sim0.3$ (similar
to that expected from a classic accretion disk) as well as an emission
line spectrum exhibiting Balmer lines and HeI and a F/G stellar spectrum.
The presence of eclipses verifies that the inclination $i\sim90^\circ$.
After correcting for
reddening  the optical luminosity is $\sim0.01$ times the X-ray power.
The intensity is comparable with that expected from a wind of the
strength that we have had to posit to account for the radio jet
collimation with a filling factor of order unity and the line widths
are compatible with those expected from a wind with $V_{W8}\sim1$.

The radio outbursts observed seem (Tingay et al. 1995; Harmon et
al. 1995) to follow the hard X-ray bursts with a lag of a
few days to a few weeks; the radio flux rises as
the X-ray flux falls.  The flattening of the hard X-ray spectrum during this
stage might be attributed to a beamed component produced in the jet.
The characteristic time delay is comparable with the wind travel time
to the radius at which the radio jet becomes optically thin.
\sec{5. Future Observational Tests}
The existence of collimated relativistic outflows in these two Galactic
superluminal sources strongly motivates a search for other examples,
particularly in known X-ray transients (HR95).
(Observations in the week
following X-ray outbursts are particularly relevant in view of the reported
behavior in  GR0~1655-40.) The two Galactic $\gamma$-ray sources, 1E1740.7-2942
and GRS 1758-258 (Mirabel 1994, Chen, Gehrels \& Leventhal 1994)
show many similarities to the two sources considered above (radio jets,
hard X-rays, upper limits on the masses of the stellar companions) as well as
some differences (detected $\gamma$-rays, correlation between the
radio and X-ray fluxes). It is also of interest to re-examine CygX-3
from which Strom et al. (1988) report mildly relativistically moving
radio components.

The counterparts
of the giant radio lobes by which the extragalactic counterparts
of these sources were
first recognized (or alternatively of W50)
should also be sought.  This search should be
most profitably carried out at
low frequency and as in the extragalactic case the
radio source sizes may be very
large ($\go10'$), depending upon the history and local gas density.
As extragalactic observations also emphasize, there are possible
strong selection effects and GRS~1915+105, GRO~1655-40 may be just
examples of a much larger class, most of which contain jets
with larger Lorentz factors and rendered invisible by beaming away
from us. These too may be found by wide field, low frequency
radio observations of other X-ray transients.

Also drawing upon the extragalactic analogy, a search for $\sim1$~GeV
$\gamma$-rays using EGRET is well-motivated and a successful detection
would strengthen the case for $e^+-e^-$ jets. Conversely, the detection
of optical or infrared Doppler-shifted emission lines would strengthen
the association with SS433 and argue for an $e-p$ jet.

Given the large inclination
derived from the radio source kinematics ($i\sim84^\circ$, HR95),
should a measurement could then be translated into an
estimate of the size of the Roche lobe of the companion star.
Understanding the size of the orbit and consequently of the accretion
disk should also help define the physical conditions
in the bipolar wind that we have invoked to account for the jet collimation.
Further constraints on the discharge in the wind can come
from observing the soft X-rays with ROSAT at energies
$\lo1$~keV during the phases when
the radio components are being formed.  If the outflow is as dense as we
propose, then the soft X-rays should be efficiently absorbed.
A quite separate bound on the wind density may be derivable by seeking
rapid variability in hard X-rays using ASCA during high states.
Observation of rapid variability on time scale on timescale $t_{var}$
would imply that any wind be optically thin the Thomson scattering beyond
a radius $r\sim ct_{var}$.  If, somewhat unexpectedly, the compact object
in either source
is a neutron star with a measurable spin period, then this test will be much
stronger.  In addition, we have suggested that the relative strength of the
emission lines from this wind should be anti-correlated with the X-ray flux.

However, perhaps the most fundamental understanding as to the nature
of these sources will come from analyzing the kinematics of the radio
components to see if there are genuine periodicities in either the timing
of the outbursts or, as tentatively supported by the observations of
HR95, the component angular velocities projected
on the sky.  It will be especially interesting to learn if, on these grounds,
the Galactic superluminal sources are more affiliated with extragalactic jets
or SS433.
\sec{Acknowledgements}  We acknowledge helpful conversations with
Charles Bailyn, Bob Hjellming, Dave Meir, Felix Mirabel and Rashid Sunyaev.
Support under NASA grant NAG 5-2757 is gratefully acknowledged.
\vfill \eject
\sec{References}

\noindent{Blandford, R.D., 92, in Active Galactic Nuclei, ed. R. Blandford,
H. Netzer, \& L. Woltjer (Berlin: Springer), p. 57

\noindent{Blandford, R. D., \& Levinson, A., 95, ApJ, 441, 79}

\noindent{Bailyn, C.D., et al., 95, Nature, 374, 701}

\noindent{Bailyn, C. D. \& Orosz, J., 95, IAU Circular No., 6173, }

\noindent{Begelman, M., et al., 80, ApJ, 238, 722}

\noindent{Begelman, M., Blandford, R.D., \& Rees, M., 84, Rev. Mod.
     Phys., 56, 255}

\noindent{Bodo, G., \& Ghisellini, G., 95, ApJl, 441, L69}

\noindent{Chen, W., Gehrels, N., \& Leventhal, M., 94, ApJ, 426, 586}

\noindent{Davidson, K., \& McCray, R., 80, ApJ, 241, 1082}

\noindent{Harmon, A., et al., 94, in Second Compton Symposium, ed. C.E.
Fichtel, N. Gehrels, \& J.P. Norris, (New York: AIP), 210}

\noindent{Harmon, A., et al., 95, Nature, 374, 703}

\noindent{Hjellming, R.M., \& Rupen, M.P., 95, Nature, 375, 464}

\noindent{Levinson, A., \& Blandford, R.D., 95, ApJ, in press}

\noindent{Meier, D., 95, preprint }

\noindent{Mirabel, I.F., 94, ApJs, 92, 369}

\noindent{Mirabel, I.F., \& Rodriguez, L.F., 94, Nature, 371, 46}

\noindent{Nath, B., 95, MNRAS, 274, 208}

\noindent{Rybicki, G. \& Lightman, A. P., 79, Radiation Processes in
Plasma, New York: Wiley}

\noindent{Strom, R. G., van Paradijs, J. \& van der Klis, M., 88,
Nature, 337, 234}

\noindent{Tingay, S.J., et al., 95, Nature, 374, 141}
\bye